\documentclass[a4paper]{article}

\usepackage{INTERSPEECH2022}

\usepackage{mathtools}
\usepackage{booktabs}
\usepackage{multirow}
\usepackage{subfig}
\usepackage{siunitx}
\newcommand{\ctext}[1]{\raise0.2ex\hbox{\textcircled{\scriptsize{#1}}}}

\newcommand{\vect}[1]{\mbox{\boldmath $#1$}}

\def\appendixautorefname~#1\null{~#1 \null}

\title{Updating Only Encoders Prevents Catastrophic Forgetting\\of End-to-End ASR Models}
\name{Yuki Takashima$^1$, Shota Horiguchi$^1$, Shinji Watanabe$^{2,3}$, Paola Garc\'{i}a$^3$, Yohei Kawaguchi$^1$}
\address{
$^1$ Hitachi, Ltd. Research \& Development Group, Japan\\
$^2$ Language Technologies Institute, Carnegie Mellon University, USA\\
$^3$ Center for Language and Speech Processing, Johns Hopkins University, USA}
\email{yuki.takashima.ot@hitachi.com}

\begin{document}

\maketitle
\begin{abstract}
In this paper, we present an incremental domain adaptation technique to prevent catastrophic forgetting for an end-to-end automatic speech recognition (ASR) model.
Conventional approaches require extra parameters of the same size as the model for optimization,
and it is difficult to apply these approaches to end-to-end ASR models because they have a huge amount of parameters.
To solve this problem, we first investigate which parts of end-to-end ASR models contribute to high accuracy in the target domain while preventing catastrophic forgetting.
We conduct experiments on incremental domain adaptation from the LibriSpeech dataset to the AMI meeting corpus with two popular end-to-end ASR models and
found that adapting only the linear layers of their encoders can prevent catastrophic forgetting.
Then, on the basis of this finding, we develop an element-wise parameter selection focused on specific layers to further reduce the number of fine-tuning parameters.
Experimental results show that our approach consistently prevents catastrophic forgetting compared to parameter selection from the whole model.
\end{abstract}

\section{Introduction}
End-to-end automatic speech recognition (ASR) has made remarkable progress in cases where a large amount of training data is available.
However, in real applications, there is often an acoustic mismatch between the training environment and the operational environment where users utilize the ASR model, and it is difficult to collect a sufficient amount
of training data in the operational environment beforehand.
Therefore, we sometimes collect target domain data during operation and continuously adapt the model using the collected data.

A typical domain adaptation method~\cite{NIPS2006_b1b0432c,mirsamadi17_interspeech} aims to improve the performance on a target domain and thus usually suffers from performance degradation on a source domain, which is known as catastrophic forgetting~\cite{journals/nn/ParisiKPKW19}.
Catastrophic forgetting becomes more severe when the model is continuously adapted to the target domain using sequentially arriving data.
To mitigate catastrophic forgetting, several incremental learning methods have been proposed~\cite{10.1109/TPAMI.2017.2773081,kirkpatrick2017overcoming,conf/iclr/YoonYLH18,conf/nips/Lopez-PazR17}.
Regularization-based methods~\cite{10.1109/TPAMI.2017.2773081,kirkpatrick2017overcoming} introduce an additional loss during adaptation to make the current model close to the original one.
Architecture-based methods~\cite{conf/iclr/YoonYLH18,rusu2016progressive} dynamically expand a model architecture for new data.
In the field of ASR, various studies~\cite{conf/asru/GhorbaniKH19,conf/interspeech/SadhuH20,9746594} have demonstrated the benefits of the incremental learning approach.
Fu~{\it et al.}~\cite{fu2021incremental} proposed an incremental learning algorithm for end-to-end ASR that uses attention distillation and knowledge distillation~\cite{hinton2015distilling} to prevent catastrophic forgetting.
Such methods require retaining the previous model~\cite{conf/iclr/YoonYLH18,fu2021incremental} or adding extra parameters of the same size as the model for optimization~\cite{rusu2016progressive,conf/asru/GhorbaniKH19}.
Therefore, adapting recent end-to-end ASR models with a huge amount of parameters is computationally expensive.
A potential solution is pruning-based domain adaptation, which reduces the number of fine-tuning parameters by pruning~\cite{DBLP:conf/naacl/GuFX21,DBLP:conf/emnlp/ZhuWZLZ021}.
In~\cite{DBLP:conf/naacl/GuFX21}, parameters of a subnetwork are fixed during adaptation to keep the performance of the source domain.
The parameters to be fixed are determined on the basis of a pruning algorithm, since the parameters selected to be pruned are considered important for processing source domain data.
However, the pruning algorithm is applied for the entire network, and there has been no investigation into the subnetwork-wise parameter freezing.
Some studies on domain adaptation of ASR models have shown that updating only a part of the layers improves the performance on the target domain~\cite{conf/interspeech/UenoMMSSYAK18,sukhadia2022domain}.
However, there has been no research on the performance against catastrophic forgetting.

In this study, we investigate which parts of the end-to-end ASR model should and should not be adapted to the target domain to prevent catastrophic forgetting during incremental domain adaptation.
For the experiments, we utilize two popular ASR models: a Transformer-based model and a recurrent neural network transducer (RNN-T).
We use LibriSpeech~\cite{conf/icassp/PanayotovCPK15} as the large-scale source domain dataset and the AMI meeting corpus~\cite{conf/mlmi/CarlettaABFGHKKKKLLLMPRW05} as the target domain dataset.
We first investigate a module-wise parameter selection to be fine-tuned and
show that adapting only the encoder can prevent catastrophic forgetting.
Then, to further reduce the number of fine-tuning parameters, we develop an element-wise parameter selection from specific layers in the model.
Experimental results demonstrate that our proposed parameter selection could select better parameters to prevent catastrophic forgetting compared to parameter selection from the entire network.

\section{End-to-end ASR architecture}
\label{sec:arch}
In this section we describe the two architectures considered in this paper and explain how the layers in each are formulated.
We explicitly define the parameters fed into each function to categorize them into the predefined layer types listed in Table~\ref{tbl:split}.
For simplicity, the bias parameters are omitted from the descriptions.

\begin{table}[tb]
\caption{Parameter splitting based on functionality. MHA denotes multi-head attention layers that include self-attention layers, source-target attention layers, or both.}
\vspace{-8pt}
\label{tbl:split}
\centering
\resizebox{\linewidth}{!}{%
\begin{tabular}{@{}lcc@{}}
\toprule
Module name & Transformer-based & RNN-T \\
\midrule
Embeddings & $\vect{\Phi}^{(\text{enc})} \cup W_\text{emb}$ &  $\vect{\Phi}^{(\text{enc})} \cup W_\text{emb}$ \\
Layer normalization & $\vect{\Psi}^{(\text{enc})} \cup \vect{\Psi}^{(\text{dec})}$ & $\vect{\Psi}'^{(\text{enc})}$ \\
MHA \& FFN in encoder & $\vect{\Theta}^{(\text{enc})}$ & -- \\
MHA \& FFN in decoder & $\vect{\Theta}^{(\text{dec})}$ & -- \\
MHA \& Conv \& FFN in encoder & -- & $\vect{\Lambda}^{(\text{enc})}$ \\
Prediction & -- & $\vect{\Lambda}^{(\text{pred})}$ \\
CTC & $W_\text{CTC}$ & -- \\
Output & $W_\text{out}$ & $\vect{\Lambda}^{(\text{joint})}$ \\
\bottomrule
\end{tabular}%
}
\end{table}

\subsection{Transformer-based model}
\label{sec:aed}

For the Transformer-based ASR model, we used a hybrid connectionist temporal classification (CTC)/attention architecture.
It consists of an encoder and decoder network.

The encoder network converts a $T_0$-length time sequence of $F$-dimensional acoustic features $X_0$ into a $T$-length sequence of $D$-dimensional hidden embeddings $E_\text{out}$, as
\begin{align}
    X &= f_\text{sub}(X_0; \vect{\Phi}^{(\text{enc})})\in\mathbb{R}^{D\times T},\label{eq:sub-samp}\\
    E &= f_\text{enc}(X; \vect{\Theta}^{(\text{enc})}, \vect{\Psi}^{(\text{enc})})\in\mathbb{R}^{D\times T}.\label{eq:enc}
\end{align}
$f_\mathsf{sub}(\cdot)$ is the sub-sampling layer including the positional encodings~\cite{karita2019comparative} parameterized by $\vect{\Phi}^{(\text{enc})}\coloneqq\vect{\Gamma} \cup \{W_0\}$, where $\vect{\Gamma}$ is the set of parameters of the convolutional sub-sampling and $W_0\in \mathbb{R}^{D\times F'}$ is the subsequent linear projection matrix.
$X$ is a sequence of intermediate representations with a length of $T(<T_0)$.
$f_\text{enc}(\cdot)$ is the $N$-stacked Transformer encoder parameterized by $\vect{\Theta}^{(\text{enc})}$ and $\vect{\Psi}^{(\text{enc})}$.
$\vect{\Theta}^{(\text{enc})}$ is a set of parameters of the linear projections in the encoders, namely,
\begin{align}
    \vect{\Theta}^{(\text{enc})}\coloneqq\bigcup_{i=1}^{N}\vect{\psi}^{(i)}_\text{self} \cup \vect{\psi}^{(i)}_\text{FFN},\label{eq:transformer_linear}
\end{align}
where $\vect{\psi}^{(i)}_\text{self}$ is a set of weight matrices for the query, key, value, and output projections of the multi-head self-attention layer in the $i$-th encoder block,
and $\vect{\psi}^{(i)}_\text{FFN}$ is a set of weight matrices of the two-layer feed-forward network~(FFN) layer in the $i$-th encoder block.
$\vect{\Psi}^{(\text{enc})}$ is a set of parameters for the element-wise affine transformation in the layer normalization layer~\cite{ba2016layer}.

The decoder network calculates the $L$-length target sequence $\left[\vect{y}_l\right]_{l=1}^{L}$ from the encoder output $E$ in a sequence-to-sequence manner, as follows:
\begin{align}
    \vect{z}_l &= f_\text{emb}(\vect{y}_l; W_{\text{emb}})\in\mathbb{R}^D,\label{eq:dec_emb}\\
    \vect{o}_l &= f_\text{dec}([\vect{z}_1, \dots, \vect{z}_{l-1}], E; \vect{\Theta}^{(\text{dec})}, \vect{\Psi}^{(\text{dec})}),\\
    \vect{\hat{y}}_l &= f_\text{out}(\vect{o}_l;W_{\text{out}}) \in\left(0,1\right)^{V}.
\end{align}
$f_\text{emb}$ is the embedding layer that converts a one-hot encoding $\vect{y}\in\{0,1\}^V$ into the corresponding embedding $\vect{z}_n\in\mathbb{R}^D$ using the projection matrix $W_{\text{emb}}\in\mathbb{R}^{D\times V}$, where $V$ is the vocabulary size.
$f_\text{dec}$ is $M$-stacked Transformer decoders parameterized by $\vect{\Theta}^{(\text{dec})}$ and $\vect{\Psi}^{(\text{dec})}$.
$\vect{\Theta}^{(\text{dec})}$ is a set of parameters of the linear projections in the decoders, denoted as
\begin{align}
    \vect{\Theta}^{(\text{dec})}\coloneqq\bigcup_{i=1}^{M} \vect{\phi}^{(i)}_\text{self} \cup \vect{\phi}^{(i)}_\text{src} \cup \vect{\phi}^{(i)}_\text{FFN},
\end{align}
where $\vect{\phi}^{(i)}_\text{self}$ and $\vect{\phi}^{(i)}_\text{src}$ are sets of parameters of the multi-head self-attention layer and multi-head source-target attention layer in the $i$-th decoder block, and
$\vect{\phi}^{(i)}_\text{FFN}$ is a set of weight matrices of two-layer FFNs in the $i$-th decoder block.
Each of $\vect{\phi}^{(i)}_\text{self}$ and $\vect{\phi}^{(i)}_\text{src}$ consists of four weight matrices, similar to $\vect{\psi}^{(i)}_\text{self}$.
A set of parameters $\vect{\Psi}^{(\text{dec})}$ consists of weight vectors for element-wise affine transformation in the layer normalization layers.
The classification layer $f_\text{out}$  converts the output from the Transformer decoder into the vocabulary-wise posterior probability.
It consists of a linear layer parameterized by $W_\text{out}\in\mathbb{R}^{V\times D}$ followed by a softmax function.

The entire network is optimized by minimizing the negative log-likelihood and the CTC loss~\cite{karita2019comparative}.
The negative log-likelihood is calculated between the target sequence and the output sequence from the decoder.
The CTC loss is determined between the target sequence and the sequence $\left[\vect{\bar{y}}_t\right]_{t=1}^{T}$, each of which is calculated from the encoder output $E = [\vect{e}_t]_{t=1}^{T}$ as follows:
\begin{align}
    \bar{\vect{y}}_t = f_\text{CTC}(\vect{e}_t; W_\text{CTC})\in(0,1)^{V\times T},
\end{align}
where $f_\text{CTC}(\cdot)$ is the CTC function using the projection matrix $W_\text{CTC}\in\mathbb{R}^{V\times D}$.
The CTC layer maps a sequence of the frame-wise label probabilities into a target label sequence to calculate the likelihood of a target sequence~(for more details, see~\cite{Graves:06icml}).

\subsection{Recurrent neural network transducer}
The RNN-T ASR model utilizes a Conformer encoder, a long short-term memory (LSTM)-based prediction network, and a joint network.
Similar to the Transformer-based model, the input acoustic features are sub-sampled before being fed into the encoder using Eq.~(\ref{eq:sub-samp}).

The encoder converts a $T$-length $D$-dimensional sequence into the same dimensional hidden embedding sequence as
\begin{align}
    [\vect{e}'_1, \dots, \vect{e}'_T] = f'_\text{enc}(X; \vect{\Lambda}^{(\text{enc})}, \vect{\Psi}'^{(\text{enc})})\in\mathbb{R}^{D\times T},
\end{align}
where $\vect{e}'_t$ is the $t$-th hidden embedding.
$f'_\text{enc}(\cdot)$ is the $N$-stacked Conformer encoder~\cite{conf/interspeech/GulatiQCPZYHWZW20} parameterized by $\vect{\Lambda}^{(\text{enc})}$ and $\vect{\Psi}'^{(\text{enc})}$.
$\vect{\Lambda}^{(\text{enc})}$ is a set of parameters of the linear projections in the encoders, namely,
\begin{align}
    \vect{\Lambda}^{(\text{enc})}\coloneqq\bigcup_{i=1}^{N} \vect{\lambda}^{(i)}_\text{pre} \cup \vect{\lambda}^{(i)}_\text{self} \cup \vect{\lambda}^{(i)}_\text{conv} \cup \vect{\lambda}^{(i)}_\text{post},
\end{align}
where $\vect{\lambda}^{(i)}_\text{self}$ is a set of parameters of the multi-head self-attention layer in the $i$-th encoder, and $\vect{\lambda}^{(i)}_\text{conv}$ is a set of parameters of convolutional kernels including pointwise and depthwise convolutions in the $i$-th encoder.
$\vect{\lambda}^{(i)}_\text{pre}$ and $\vect{\lambda}^{(i)}_\text{post}$ are sets of parameters of two-layer FFN layers before the self-attention layer and after the convolution layer, respectively.
$\vect{\Psi}$ is the set of parameters of the affine transformations in the layer normalization layers.

The prediction network $f_\text{pred}$ calculates the $L$-length hidden representations $[\vect{h}_l]_{l=1}^L$ in a sequence-to-sequence manner, as follows:
\begin{align}
    \vect{h}_l = f_\text{pred}([\vect{z}_1, \dots, \vect{z}_{l-1}]; \vect{\Lambda}^{(\text{pred})})\in\mathbb{R}^D,
\end{align}
where $\vect{\Lambda}^{(\text{pred})}$ is the parameters of LSTM~\cite{huang2020improving}.
$\vect{z}_l$ is the hidden representation calculated using Eq.~(\ref{eq:dec_emb}).

The joint network $f_\text{joint}(\cdot)$ integrates each frame of the encoder output $\vect{e}'_t$ and the output $\vect{h}_{l}$ from the prediction network using a feed-forward network as follows:
\begin{align}
    \bar{y}_{t, l} = f_\text{joint}(\vect{e}'_{t}, \vect{h}_{l}; \vect{\Lambda}^{(\text{joint})})\in\left(0,1\right)^V,
\end{align}
where $\vect{\Lambda}^{(\text{joint})}$ is
a set of weight matrices for the encoder-subspace, prediction-subspace, and subspace-output projections.

The likelihood of the target sequence is calculated as the objective function by summing the possibilities of all possible alignments for the target sequence~\cite{graves2012sequence} as well as CTC.

\section{Method}
\subsection{Module-wise parameter selection}
\label{sec:investigation}
Recent end-to-end ASR models consist of several types of layers, as mentioned in Section~\ref{sec:arch}.
These layers capture different characteristics and are typically utilized to transfer their characteristics to the target-domain model~\cite{conf/interspeech/UenoMMSSYAK18,sukhadia2022domain}.
With this observation in mind, we categorize them into eight modules, as shown in Table~\ref{tbl:split}.
We assume that each module differs in adaptability to the target domain and a forgetting property of the source domain.
Our first method, module-wise parameter selection, aims to clarify whether we can choose the optimal modules to be updated to achieve end-to-end ASR models that perform well on both source and target domains.
Given a model well-trained on the source domain dataset, we select several modules (shown in Table~\ref{tbl:split}) to be fine-tuned.
The parameters of the remaining modules are kept unchanged during adaptation.
We conducted multiple stages of adaptation using subdivided data of the target domain.
Note that each subdivided data is used only in one of the stages and is not carried over between stages.

\subsection{Element-wise parameter selection for specific layers}
Each module in Table~\ref{tbl:split} still has a large number of parameters, especially in the encoder and decoder.
To further reduce the number of fine-tuning parameters, we propose an element-wise parameter selection for specific modules.
Given a model trained on the source domain, we first select a subset of parameters within specific modules.
In this work, we determine these specific modules on the basis of the investigation in Section~\ref{sec:investigation} to prevent catastrophic forgetting.
For parameter selection, we utilize random selection and magnitude-based selection, the latter of which is commonly used in network pruning~\cite{DBLP:conf/emnlp/ZhuWZLZ021,han2015learning}.
Then, we fine-tune only the selected parameters during adaptation.

\section{Experiments}
\subsection{Conditions}

\begin{table}[tb]
\caption{Statistics of adaptation, development, and test sets.}
\vspace{-8pt}
\label{tbl:dataset}
\centering
\resizebox{\linewidth}{!}{%
\begin{tabular}{@{}llcc@{}}
\toprule
& &\#Utterances & Duration (h) \\
\midrule
Adaptation sets & \:AMI-stage1 & 25,904 & 18.3 \\
&\:AMI-stage2 & 28,746 & 19.7 \\
&\:AMI-stage3 & 24,265 & 18.0 \\
\midrule
Development set & \:AMI & 13,059 & 8.9 \\
\midrule
Test sets &\:LibriSpeech-clean & 2,620 & 5.4 \\
&\:LibriSpeech-other & 2,939 & 5.3 \\
&\:AMI-scenario & 7,503 & 5.7 \\
&\:AMI-non-scenario & 5,109 & 3.0 \\
\bottomrule
\end{tabular}%
}
\end{table}

We conducted incremental domain adaptation experiments using the LibriSpeech dataset~\cite{conf/icassp/PanayotovCPK15} as the source domain and the AMI meeting corpus~\cite{conf/mlmi/CarlettaABFGHKKKKLLLMPRW05} as the target domain.
For the AMI corpus, we selected the official Full-corpus-ASR partition and randomly divided the scenario meetings in the training set into three subsets to conduct three stages of adaptation.
AMI-stage\{1,2,3\} were used in this order.
We also divided the test set into two subsets, scenario meetings and non-scenario meetings, which were used to evaluate the performance on target domain data and out-of-domain data, respectively.
Additionally, we used the test sets of the LibriSpeech dataset to measure performance degradation on the source domain.
The statistics of each dataset are listed in Table~\ref{tbl:dataset}.

We used 80-dimensional log-mel features with a \SI{25}{\ms} frame length and \SI{10}{\ms} frame shift as input for the end-to-end ASR models.
We used a 5k vocabulary based on the subword segmentation algorithm~\cite{conf/acl/Kudo18} with 5,002 output dimensions $V$ including subword tokens.
For adaptation, we utilized the Adam optimizer~\cite{journals/corr/KingmaB14} with a fixed learning rate of $1\times10^{-3}$.
For decoding, we used a beam width of 10.

We used the pre-trained Transformer-based model with roughly 78.4M parameters in ESPnet~\cite{conf/interspeech/WatanabeHKHNUSH18}, which has 12 Transformer encoder blocks and six Transformer decoder blocks.
The number of attention heads was eight.
The sub-sampling factor and the embedding dimensionality were set to 4 ({\it i.e.}, $T={T_0}/4$) and $D=512$.
For the convolutional sub-sampling layer, $\vect{\Phi}^{(\text{enc})}$ has two convolution layers
with a $3\times 3$ filter followed by a ReLU activation with a channel size of 512.
The model was optimized by minimizing the negative log-likelihood and CTC loss with an interpolation weight $0.3$.
For inference, we implemented two decoding methods: joint CTC/attention decoding and CTC decoding.

For the RNN-T model, we also used the pre-trained model with roughly 88.7M parameters in ESPnet, which has 12 Conformer encoder blocks with four attention heads.
The embedding dimensionality was also set to $D=512$.
For the convolutional sub-sampling layer, $\vect{\Phi}^{(\text{enc})}$ has four convolution layers
with a $3\times 3$ filter followed by a ReLU activation with channel sizes 64, 64, 128, 128.
The prediction network has an embedding layer of 1,024 units and one LSTM layer of 512 units.
The joint network is composed of FFNs with 512 hidden units.

For the investigation of the optimal adapting parameters of the model, we split the parameters of the Transformer-based model and the RNN-T model into six and five modules, respectively, as shown in Table~\ref{tbl:split}.

\subsection{Results}

\begin{figure*}[t]
    \centering
    \subfloat[][Hybrid CTC/attention decoding]{\includegraphics[width=0.31\linewidth]{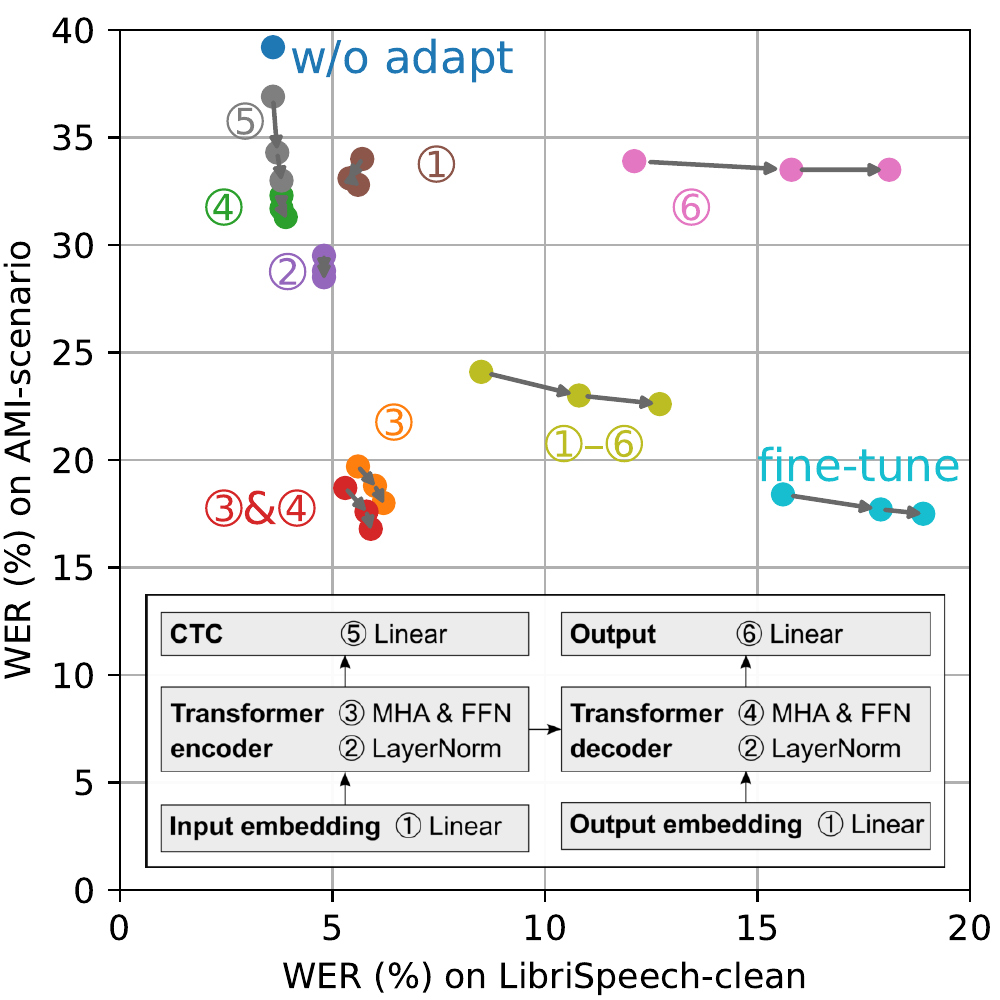}\label{fig:res1-Transformer}}
    \hfill
    \subfloat[][CTC decoding]{\includegraphics[width=0.31\linewidth]{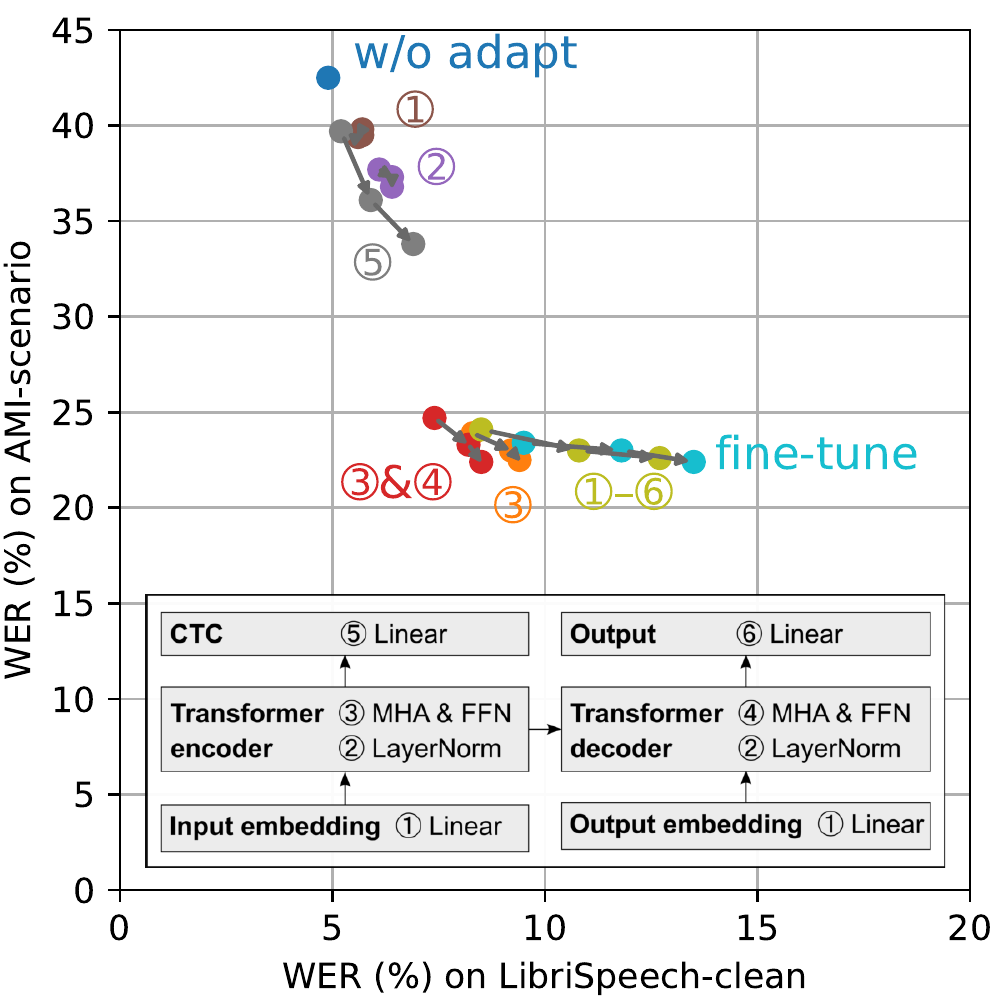}\label{fig:res1-ctc}}
    \hfill
    \subfloat[RNN-T]{\includegraphics[width=0.31\linewidth]{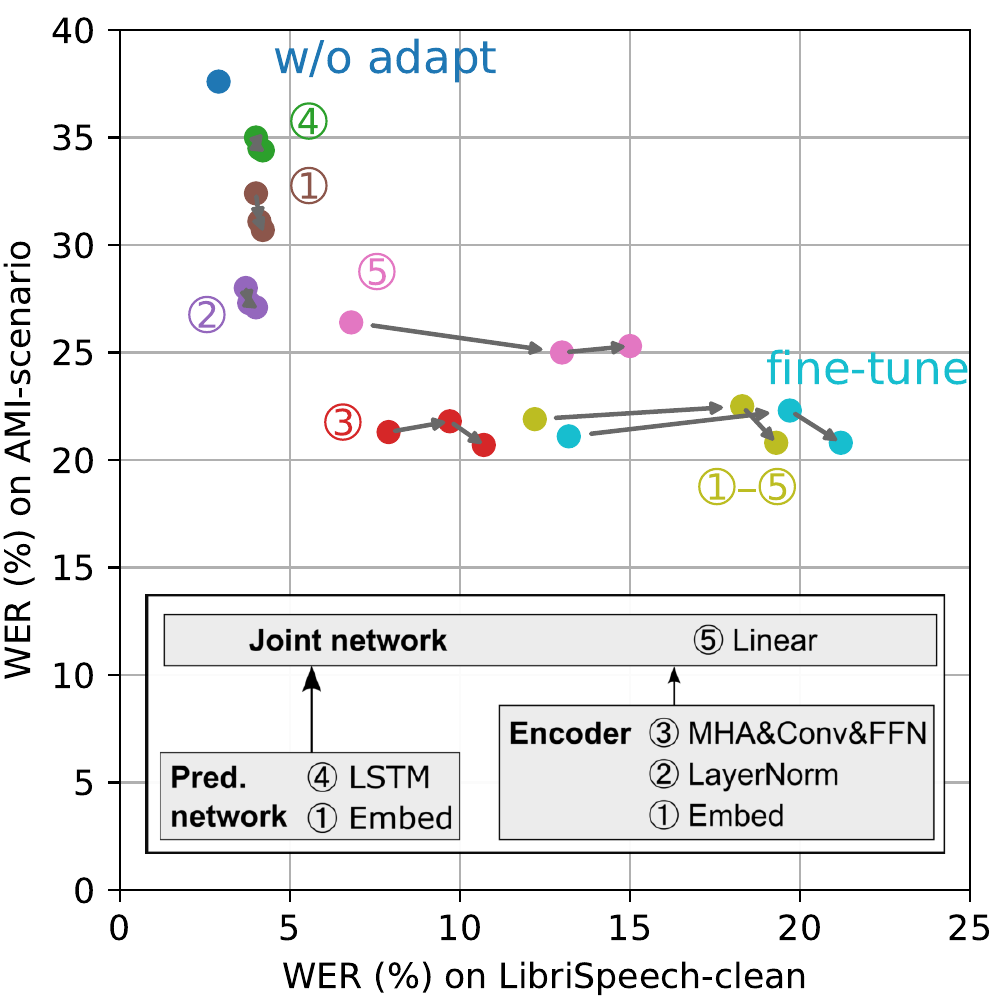}\label{fig:res1-rnnt}}
    \vspace{-6pt}
    \caption{WERs (\%) with different adapting modules. (a) and (b) show the results using different decoding methods in the Transformer-based model. (c) shows the results of the RNN-T model. The arrows indicate the trajectory of adaptation.}
    \label{fig:res1}
    \vspace{-10pt}
\end{figure*}

Figure~\ref{fig:res1} shows the word error rates~(WERs) for the Transformer-based model and the RNN-T model.
In this figure, the bottom left is better, and going right means catastrophic forgetting.
Fine-tuning all the parameters (`fine-tune') caused catastrophic forgetting.
As shown in Fig.~\ref{fig:res1-Transformer}, fine-tuning weight matrices in the Transformer (\ctext{3}\&\ctext{4}) achieved the best performance on both domains.
In particular, weight matrices in the encoder (\ctext{3}) contributed to improving the performance on the target domain while those in the decoder (\ctext{4}) contributed to preventing catastrophic forgetting.
On the other hand, fine-tuning the output layer (\ctext{6}) caused severe catastrophic forgetting even though this layer has only 3.3\% of the model parameters.
We can see that modules related to language characteristics (except for the output module) tend to retain the performance of the source domain.
We hypothesize that the model could have learned sufficient language characteristics from the LibriSpeech dataset.
Moreover, the embedding module achieved a small improvement on the target domain compared to other modules as well as the previous work~\cite{DBLP:conf/emnlp/ZhuWZLZ021}, in which the embedding layer did not yield consistent differences for adaptation.
Figure~\ref{fig:res1-ctc} shows the results of CTC decoding, where the results of fine-tuning the decoder (\ctext{4}) and the output layer (\ctext{6}) are not plotted because these results are identical to the result without adaptation.
Here, we can see similar trends with CTC/attention decoding.

In the case of the RNN-T model shown in Fig.~\ref{fig:res1-rnnt}, we also see similar trends to the Transformer-based model.
Fine-tuning the joint network (\ctext{4}) caused catastrophic forgetting with poor performance on the target domain.
Fine-tuning the Conformer encoder (\ctext{3}) achieved competitive WERs with fine-tuning all the parameters on the target domain while keeping degradation on the source domain small.
From these results, we conclude that the encoder in the end-to-end ASR model is an optimal choice for adaptation.
The encoder is a counterpart of the acoustic model, and our conclusion is in good agreement with the well-known traditional acoustic model adaptation~\cite{sagayama2001analytic}.
We also conclude that adaptation of the encoder contributes to the prevention of catastrophic forgetting.

\begin{table}[tb]
\caption{WERs (\%) for different methods on four test sets. \#Params denotes the number of fine-tuning parameters. `\ctext{3}\&\ctext{4}' corresponds to the numbers shown in Fig.~\ref{fig:res1}.}
\vspace{-8pt}
\label{tbl:res_comp}
\centering
\scalebox{0.9}{
\setlength{\tabcolsep}{3.0pt}
\begin{tabular}{@{}llcccc@{}}
\toprule
Test set & Method & \#Params & Stage1 & Stage2 & Stage3 \\
\midrule
LibriSpeech- & EWC & 78.4M  & 5.7 & 6.0 & 6.0 \\
clean & Fine-tuning \ctext{3}\&\ctext{4} & 62.9M & 5.3 & 5.8 & 5.9 \\
\midrule
LibriSpeech- & EWC & 78.4M  & 13.5 & 14.6 & 15.0 \\
other & Fine-tuning \ctext{3}\&\ctext{4} & 62.9M  & 12.5 & 13.9 & 14.4 \\
\midrule
AMI- & EWC & 78.4M  & 18.4 & 17.7 & 17.4 \\
scenario & Fine-tuning \ctext{3}\&\ctext{4} & 62.9M  & 18.7 & 17.6 & 16.8 \\
\midrule
AMI- & EWC & 78.4M  & 29.5 & 28.8 & 28.4 \\
non-scenario & Fine-tuning \ctext{3}\&\ctext{4} & 62.9M  & 31.0 & 29.1 & 27.9 \\
\bottomrule
\end{tabular}
}
\vspace{-15pt}
\end{table}

Next, we conducted comparative experiments with our approach and the conventional incremental learning, EWC~\cite{kirkpatrick2017overcoming}.
We utilized the Transformer-based model with hybrid CTC/attention decoding shown in Fig.~\ref{fig:res1-Transformer}.
As shown in Table~\ref{tbl:res_comp}, our approach outperformed EWC on all test sets in the final stage.
However, in stage 1 and 2, the WERs of our approach on AMI-non-scenario were slightly higher than those of EWC.
We hypothesize that EWC was able to adapt the model to the AMI dataset well in the early stages of adaptation because it fine-tuned all the parameters of the model.
Our approach has the advantage of not requiring any pre-computation or additional parameters, while EWC requires the Fisher information matrix calculated on the source domain.
These results demonstrate that the simple fine-tuning approach is effective for incremental domain adaptation.

\begin{figure}[t]
\vspace{-3pt}
    \centering
        \includegraphics[clip,keepaspectratio, width=\linewidth]{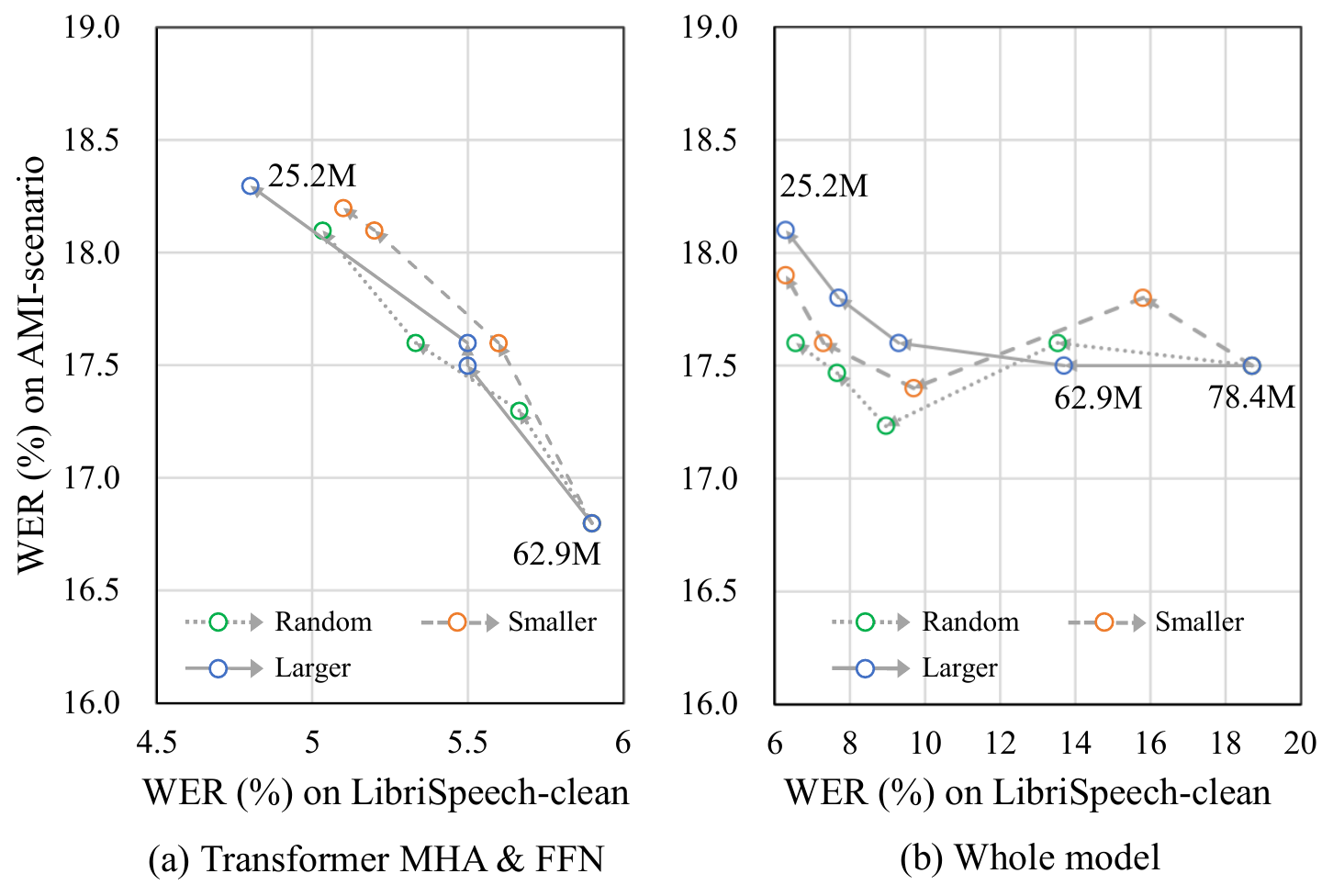}
        \vspace{-16pt}
        \caption{WERs (\%) at the third stage of adaptation with various numbers of adapting parameters: 62.9M, 50.3M, 37.8M, and 25.2M. 
        62.9M corresponds to the number of parameters of Transformer MHA \& FFN. The WERs of the random selection are mean values across three trials.}
        \label{fig:res_mask}
        \vspace{-15pt}
\end{figure}

Finally, we show the results of the element-wise parameter selection.
We again utilized the Transformer-based model with hybrid CTC/attention decoding.
We evaluated three parameter selection approaches: random, smaller-magnitude, and larger-magnitude.
We applied parameter selection to weight matrices in the Transformer and the whole model that correspond to \ctext{3}\&\ctext{4} and `fine-tune' in Fig.~\ref{fig:res1}.
As shown in Fig.~\ref{fig:res_mask}, parameter selection with only 37.8M parameters achieved competitive results on the target domain compared with fine-tuning all the parameters while preventing catastrophic forgetting.
However, we could not observe a large difference among different parameter selections.
In the case of the whole model, we obtained inconsistent WERs among different numbers of fine-tuning parameters, except for larger-magnitude selection.
These results show that fine-tuning a part of the parameters within the Transformer achieves a consistent performance regardless of the difference in the parameter selection.

\section{Conclusions}
In this paper, we presented a simple but effective incremental domain adaptation strategy to prevent catastrophic forgetting for end-to-end ASR models.
Our experiments revealed that parameter selection can sufficiently prevent catastrophic forgetting without the complex procedures used in the standard incremental domain adaptation methods~\cite{10.1109/TPAMI.2017.2773081,kirkpatrick2017overcoming}.
In future work, we will apply our findings to conventional incremental domain adaptation methods.

\bibliographystyle{IEEEtran}

\bibliography{mybib}

\end{document}